\documentclass[twocolumn,aps,prl,floatfix,showpacs,superscriptaddress]{revtex4}

\usepackage{graphicx}
\usepackage{dcolumn}
\usepackage{bm}

\def\simle{\hspace*{0.2em}\raisebox{0.5ex}{$>$}
     \hspace{-0.8em}\raisebox{-0.3em}{$\sim$}\hspace*{0.2em}} \def\simle{\hspace*{0.2em}\raisebox{0.5ex}{$<$}
    \hspace{-0.8em}\raisebox{-0.3em}{$\sim$}\hspace*{0.2em}}

\def\s2w{\sin^2\theta_W}

\begin{document}
\preprint{Draft}

\title{The Strange Quark Contribution to the Proton's Magnetic Moment}

\author{D.~T.~Spayde}
\affiliation{Department of Physics, University of Illinois at
  Urbana-Champaign, Urbana, IL 61801}

\author{T.~Averett}
\affiliation{Department of Physics, College of William and Mary,
  Williamsburg, VA 23187}

\author{D.~Barkhuff}
\affiliation{Bates Linear Accelerator Center, Laboratory for Nuclear
  Science, Massachusetts Institute of Technology, Middleton,
  MA 01949}

\author{D.~H.~Beck}
\affiliation{Department of Physics, University of Illinois at
  Urbana-Champaign, Urbana, IL 61801}

\author{E.~J.~Beise}
\email{beise@physics.umd.edu}
\affiliation{Department of Physics, University of Maryland, College
  Park, MD 20742}

\author{H.~Breuer}
\affiliation{Department of Physics, University of Maryland, College
  Park, MD 20742}

\author{R.~Carr}
\affiliation {W.K.Kellogg Radiation Laboratory, California Institute
  of Technology Pasadena, CA 91125}

\author{S.~Covrig}
\affiliation {W.K.Kellogg Radiation Laboratory, California Institute
  of Technology Pasadena, CA 91125}

\author{G.~Dodson}
\affiliation{Bates Linear Accelerator Center, Laboratory for Nuclear
  Science, Massachusetts Institute of Technology, Middleton,
  MA 01949}

\author{K.~Dow}
\affiliation{Bates Linear Accelerator Center, Laboratory for Nuclear
  Science, Massachusetts Institute of Technology, Middleton,
  MA 01949}

\author{M.~Farkhondeh}
\affiliation{Bates Linear Accelerator Center, Laboratory for Nuclear
  Science, Massachusetts Institute of Technology, Middleton,
  MA 01949}

\author{B.~W.~Filippone}
\affiliation {W.K.Kellogg Radiation Laboratory, California Institute
  of Technology Pasadena, CA 91125}

\author{W.~Franklin}
\affiliation{Bates Linear Accelerator Center, Laboratory for Nuclear
  Science, Massachusetts Institute of Technology, Middleton,
  MA 01949}

\author{R.~Hasty}
\affiliation{Department of Physics, University of Illinois at
  Urbana-Champaign, Urbana, IL 61801}

\author{T.~M.~Ito}
\affiliation {W.K.Kellogg Radiation Laboratory, California Institute
  of Technology Pasadena, CA 91125}

\author{W.~Korsch}
\affiliation{Department of Physics and Astronomy, University of
  Kentucky, Lexington, KY 40506}

\author{S.~Kowalski}
\affiliation{Laboratory for Nuclear Science, Massachusetts Institute
  of Technology, Cambridge, MA 02139}

\author{R.~D.~McKeown}
\affiliation {W.K.Kellogg Radiation Laboratory, California Institute
  of Technology Pasadena, CA 91125}

\author{B.~Mueller}
\affiliation{Physics Division, Argonne National Laboratory, Argonne,
  IL 60439}

\author{M.~L.~Pitt}
\affiliation{Department of Physics, Virginia Polytechnic Institute and
  State University, Blacksburg,  VA 24061}

\author{M.J.~Ramsey-Musolf}
\affiliation {W.K.Kellogg Radiation Laboratory, California Institute
  of Technology Pasadena, CA 91125}
\affiliation{Department of Physics, University of Connecticut, Storrs, CT  06269}

\author{J.~Ritter}
\affiliation {W.K.Kellogg Radiation Laboratory, California Institute
  of Technology Pasadena, CA 91125}
\affiliation{Department of Physics, Virginia Polytechnic Institute and
  State University, Blacksburg,  VA 24061}

\author{R.~Tieulent}
\affiliation{Department of Physics, University of Maryland, College
  Park, MD 20742}

\author{E.~Tsentalovich}
\affiliation{Bates Linear Accelerator Center, Laboratory for Nuclear
  Science, Massachusetts Institute of Technology, Middleton,
  MA 01949}

\author{S.~P.~Wells}
\affiliation{Department of Physics, Louisiana Tech University, Ruston,
  LA 71272}

\author{B.~Yang}
\affiliation{Bates Linear Accelerator Center, Laboratory for Nuclear
  Science, Massachusetts Institute of Technology, Middleton,
  MA 01949}

\author{T.~Zwart}
\affiliation{Bates Linear Accelerator Center, Laboratory for Nuclear
  Science, Massachusetts Institute of Technology, Middleton,
  MA 01949}

\date{\today}

\begin{abstract}
We report a new determination of the strange quark contribution to the
proton's magnetic form factor at a four-momentum transfer $Q^2$ = 0.1(GeV/c)$^2$
from parity-violating $e$-$p$ elastic scattering. The result uses a revised
analysis of data from the SAMPLE experiment which was carried out
at the MIT-Bates Laboratory. The data are
combined with a calculation of the proton's axial form factor $G_A^e$ to determine
the strange form factor $G_M^s$($Q^2$=0.1)=0.37$\pm$0.20$\pm$0.26$\pm$0.07. The
extrapolation of $G_M^s$ to its $Q^2=0$ limit and comparison with calculations
is also discussed.

\end{abstract}


\maketitle

In 1988, Kaplan and Manohar~\cite{Kap88} showed that unique
information about the contribution of sea quarks to ground state nucleon
properties, such as its spin, charge and magnetic moment, can be
gained through study of neutral weak probes of the nucleon such as neutrino-nucleon
scattering. Soon after, McKeown~\cite{BMcK89} and
Beck~\cite{Bec89} outlined a possible program of experiments in
parity-violating electron scattering that, when combined with existing
measurements of nucleon electromagnetic form factors, would
allow the identification of possible strange quark contributions
to the proton's charge and magnetism. Experiments have since been
carried out~\cite{Mue97,Spa00,Has00,Ani00,Maa03}
and proposed~\cite{Bec00,Kum99} at several electron scattering facilities,
and many theoretical calculations of the strange quark components of these
neutral weak matrix elements have appeared in the literature.

In this paper we present a new determination of the strange quark contribution
to the proton's magnetic form factor, $G_M^s$, from the parity-violating
$e$-$p$ elastic scattering data of the SAMPLE experiment, which
was carried out at the MIT-Bates Laboratory in 1998. Preliminary results
were presented in~\cite{Spa00}: in the
intervening period two measurements with a deuterium
target were performed along with a more detailed analysis of the hydrogen
data.  The deuterium data are relatively insensitive to the strange vector
matrix elements, but provide the first experimental
information about the nucleon's neutral weak axial form factor as seen by an
electron probe, $G_A^e$, which is necessary for a reliable extraction of $G_M^s$.  
Preliminary results using the first of the two
deuterium measurements~\cite{Has00} indicated a relatively small
contribution from strange quarks to the proton's magnetic moment
but an unexpectedly large deviation of the axial form factor from
theoretical expectation as computed in~\cite{Zhu00}. These data have been further
analyzed, and a new experiment was carried out at lower momentum transfer,
and both measurements are now in good agreement with the calculation.
The deuterium data are discussed in a separate report~\cite{Ito03}.
Here we re-evaluate the implications of the hydrogen data in
light of this new conclusion, using the calculation of~\cite{Zhu00} for
the axial form factor. We also assess its implication for the
contribution of strange quarks to the proton's magnetic moment using a theoretical
extrapolation of $G_M^s$ to its static limit.

In parity-violating elastic $e$-$p$ scattering, the asymmetry in the scattering
cross section with respect to the incident electron's helicity is
\begin{eqnarray}
\label{eq:pvee}
& A_{PV} = \frac{d\sigma_R - d\sigma_L}{d\sigma_R + d\sigma_L}
 = -\frac{G_FQ^2}{4\pi\alpha\sqrt{2}}
\times  \left[\frac{1}{\varepsilon \left(G_E^\gamma\right)^2 +
\tau \left(G_M^\gamma \right)^2}\right] \times  \nonumber \\
& \left[ \varepsilon G_E^Z G_E^\gamma + \tau G_M^Z G_M^\gamma
- \varepsilon^\prime\left(1-4\s2w\right) G_A^e G_M^\gamma \right]
\end{eqnarray}
where $\varepsilon=\left[1+2(1+\tau)\tan\frac{\theta}{2}\right]^{-1}$,
$\tau=Q^2/4M_p^2$, and
$\varepsilon^\prime = \sqrt{\tau\left(1+\tau\right)\left(1-\varepsilon^2\right)}$.
At the backward-angle kinematics of the SAMPLE experiment, the $e$-$p$ asymmetry is
dominated by the contribution from the magnetic neutral weak form factor, $G_M^Z$. When
combined with measurements of the proton and neutron electromagnetic form factors,
the strange quark component can be extracted through the relation
\begin{equation}
\label{eq:gms}
G_M^s = \left( 1-4\s2w \right)\left(1+R_V^p\right)G_M^p -\left(1+R_V^n\right)G_M^n
- G_M^Z \, .
\end{equation}
The radiative corrections $R_V^{p,n}$~\cite{MusR94} represent (small)
contributions from higher order processes. While the SAMPLE measurement
is dominated by this term, it is also sensitive to the proton's neutral weak
axial form factor $G_A^e$. At tree level,  isospin symmetry relates $G_A^e$ to the
axial vector coupling that enters neutron decay. Small corrections of order ($-10$)\%
are generated by strange quarks~\cite{FiJi01}. Electroweak radiative corrections introduce
a more substantial, ${\cal O}(-50\%)$ effect. While electroweak corrections that renormalize
the individual quark axial vector currents can be reliably computed, other effects that
involve strong and weak $qq$ correlations, such as $Z-\gamma$ box graphs and the nucleon
``anapole moment''~\cite{Zel57}, present a theoretical challenge. Predictions for these
corrections, including estimates of the theoretical uncertainties obtained
with chiral perturbation theory ($\chi$PT), were obtained in~\cite{Mus90} and updated
in~\cite{Zhu00}. The results of the SAMPLE deuterium measurements are consistent with
these predictions, and in what follows,
we use the value for $G_A^e$ obtained in~\cite{Zhu00} to interpret the results
of the SAMPLE $e$-$p$ measurement.

The SAMPLE $e$-$p$ measurement was carried out at the MIT-Bates Linear Accelerator
Center in 1998. A beam of 200~MeV circularly polarized electrons was incident on a
40~cm liquid hydrogen target, and
{\v C}erenkov light from backward-scattered electrons were detected in an
array of ten photomultiplier tubes after reflection from ellipsoidal mirrors.
The yield in each photomultiplier tube was integrated over the 25~$\mu$sec
long beam pulse and sorted by beam helicity state. The measured asymmetry was computed
as the difference between the yields in the two helicity states over the sum, with
corrections coming from helicity correlations in the beam, and from dilution factors
associated with the beam polarization (36.2$\pm$0.1\%), electromagnetic
radiative corrections, and electromagnetic background ($\sim$30\% of the yield).
Preliminary results were published in~\cite{Spa00}, in which additional details
of the experimental method can be found.

Subsequent refinement of the data analysis, along with development
of a {\sc GEANT}-based Monte Carlo simulation~\cite{GEANT} of the full experimental
geometry, has revealed three corrections, all of which act to increase the
magnitude of the experimental asymmetry.  First, the electromagnetic radiative
corrections were recomputed within the context of the simulation, whereas
in~\cite{Spa00} they were computed at the central kinematics of each detector,
resulting in a 4\% increase in the dilution factor. In both cases
a spin-dependent~\cite{Kuc83} modification to the formalism of Mo and
Tsai~\cite{Mo69} was used to compute the radiative effects. In the simulation,
scattered electron events were generated uniformly in energy, angle and
along the length of the 40~cm target. Energy loss due to ionization and collisions
in the aluminum entrance window to the target, and in the thickness of
liquid hydrogen upstream of the randomly chosen interaction point,
was accounted for before computation of the scattered electron
kinematics. Each scattered electron was assigned a cross section
and a parity-violating asymmetry, and propagated through the target
exit windows and the scattering chamber. A detection efficiency based on
the velocity of the outgoing electron and the path length of the event's
track in the {\v C}erenkov medium was combined with the computed cross
section as an event weight. The radiative correction factor of approximately
1.13 was evaluated separately for each detector module, and was computed as
the ratio of the (weighted) asymmetry without and with the radiative effects included.

Secondly, a background associated with threshold photo-pion production,
which had been neglected in~\cite{Spa00}, was evaluated using the {\sc GEANT}
simulation. Such processes contribute to the detector yield through
their decay products, but have a negligible parity-violating
asymmetry~\cite{Che01}. The $\pi^0$ ($\pi^+$) channel was modeled based
on data from~\cite{Ber98} (\cite{Kor99}). The $\pi^+$ production yield
was found to be consistent with experimental observation of an exponential
tail in the detector signal corresponding to the arrival of decay products of
secondary muons. The net additional dilution factor coming from the pion
background was 1.04.

The third modification to the previous analysis was in the treatment of the
background coming from charged particles that were not blocked by shutters
placed in front of the photomultiplier tubes. As discussed in~\cite{Spa00},
the net measured ``shutter closed'' background asymmetry was
consistent with zero,  but the detector-by-detector distribution appeared to have a
nonstatistical component, and a systematic error accounting for the
nonstatistical behavior was added to the experimental uncertainty for each
individual detector before combining them. Subsequent analysis revealed
that this shutter closed distribution  had a $\phi$-dependence which fit the function
$f(\phi )=A_0 + A_1\cos{(2\phi + \phi_0)}$ with a significantly better $\chi^2$/d.o.f
(8.0/7) than the presumed flat distribution (33.3/9). The OPEN shutter data
did not show such behavior with statistical significance. The $A_0$ coefficient,
$-$0.06$\pm$0.71~ppm after all dilution corrections, was then subtracted
from the OPEN asymmetry and its uncertainty added in quadrature.
This method produced a 5\% larger result than the method used in~\cite{Spa00}.
The three effects combined result in an experimental $e$-$p$ parity-violating 
asymmetry of~\cite{Spa01}
\begin{equation}
A(Q^2=0.1) = -5.61\pm 0.67 \pm 0.88\, {\mathrm {ppm}}\, .
\end{equation}
The Monte Carlo simulation was also used to determine the appropriate theoretical
asymmetry to which the data should be compared. Averaging over detector and target length
acceptance effects results in an approximately 3\% smaller theoretical asymmetry,
\begin{equation}
A(Q^2=0.1) = -5.56 + 3.37 G_M^s + 1.54 {G_A^e}^{(T=1)}~{\mathrm {ppm}} \, .
\end{equation}
The small isoscalar component of $G_A^e$ has been absorbed into the first term.
In the model, dipole form factors were used for $G_{E,M}^p$ and $G_M^n$, and the
Galster parameterization~\cite{Gal71} for $G_E^n$.

\begin{figure}
\begin{center}
\includegraphics[width=3.3in]{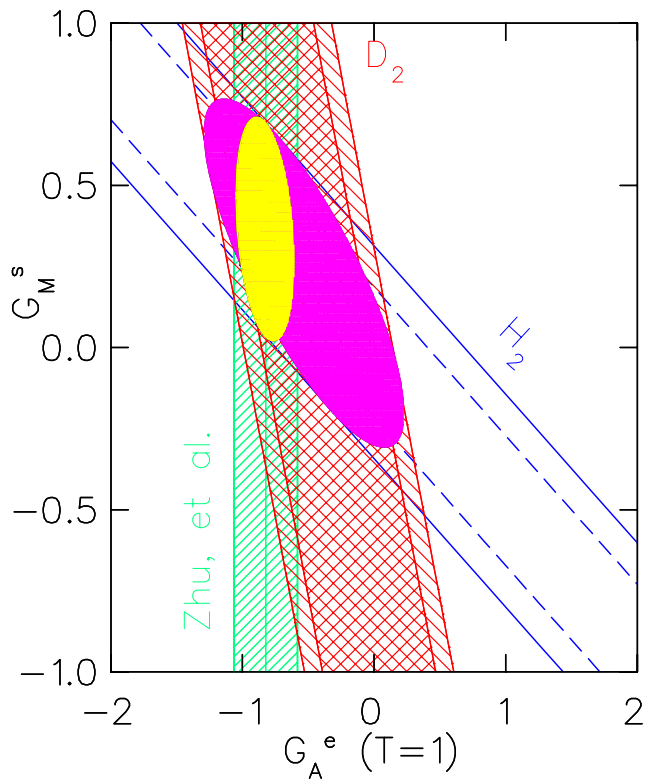}
\caption{Uncertainty bands of $G_M^s$ vs.~${G_A^e}^{(T=1)}$ at $Q^2$=0.1 (GeV/c)$^2$
resulting from the two 200~MeV data sets of the SAMPLE experiment. Also
shown is the uncertainty band of the theoretical expectation
${G_A^e}^{(T=1)}$ as computed by~\protect{\cite{Zhu00}} extrapolated to the
same momentum transfer. The smaller ellipse corresponds to the 1~$\sigma$ overlap of the
hydrogen data and the theoretical prediction, the larger one the 1-$\sigma$ overlap of
the two data sets.}
\label{fig:gms-figa}
\end{center}
\end{figure}

The two SAMPLE deuterium measurements were also analyzed using the {\sc GEANT}
simulation, as described in~\cite{Ito03}. Both measurements are in agreement with the
theoretical prediction for the asymmetry using the electroweak radiative corrections
of~\cite{Zhu00} in the computation of ${G_A^e}^{(T=1)}(Q^2)$. The results for the
two 200~MeV data sets, along with the computation of~\cite{Zhu00}, are shown in
Figure~\ref{fig:gms-figa} as 1-$\sigma$ bands in the space of
$G_M^s$ vs.~${G_A^e}^{(T=1)}$. Both the overlap of the hydrogen data and the calculation,
and the overlap of the two data sets, are shown as ellipses (1-$\sigma$), demonstrating
the good agreement between the deuterium data and the theoretical expectation
of ${G_A^e}^{(T=1)}=-0.83\pm 0.26$.  Using this value results in
\begin{equation}
\label{eq:gmsresult}
G_M^s(Q^2=0.1) = 0.37 \pm 0.20 \pm 0.26 \pm 0.07
\end{equation}
where the uncertainties are statistical, experimental systematic and the uncertainty
due to electroweak radiative corrections, respectively. Inclusion of the SAMPLE deuterium
data in the extraction of $G_M^s$ produces essentially the same result. Our new result
is slightly shifted in the positive direction relative to, but consistent with, 
the analysis in Ref.~\cite{Has00}.

The SAMPLE results for $G_M^s$, together with the considerable body of theoretical work on this topic
undertaken over the past decade, has substantially sharpened our picture of strange quark contributions
to the nucleon's electromagnetic structure. Here, we summarize the insights provided by the various
theoretical studies of strange magnetism (for a more extensive review, see~\cite{Bec01}).

The most direct, first principles approach in QCD relies on lattice
simulations. Such calculations are particularly challenging for operators such as
${\bar s}\gamma_\mu s$
that do not connect with nucleon sources, since one must sum disconnected quark loops over all
lattice sites. Although a complete calculation is prohibitively time consuming and expensive,
numerically tractable estimates can be obtained by approximating the sum with so-called
$Z_2$ noise methods \cite{Dong:1993pk}. A few calculations have been carried out, but with little consensus on the value of $G_M^s(Q^2)$.
The calculation in~\cite{Don98} that relied on a relatively small
sample of gauge field configurations and large number of $Z_2$ noise vectors yielded a
statistically significant, non-zero, negative value for $G_M^s$ over a range of momentum
transfer, with $\mu_s=G_M^s(0)=-0.36\pm0.20$.
In contrast, the computation reported in~\cite{Lew03}, which drew upon a much larger gauge
configuration sample and smaller sample of noise vectors, found $\mu_s=0.05\pm 0.06$.
A lattice-based computation by the Adelaide group is similarly
consistent with zero~\cite{Lei00},
and our experimental result tends to favor  these more recent calculations.
One anticipates that further refinements in lattice methods, such as the use of
chiral fermions and partially-quenched or unquenched computations, will eventually
lead to a consensus for the precise value of $G_M^s$.

Alternative treatments  have been pursued using dispersion relations and chiral perturbation
theory ($\chi$PT). In principle, these methods are model-independent since they
rely on general properties of QCD, such as analyticity and
causality (dispersion relations) or the approximate chiral symmetry of QCD ($\chi$PT) to
organize in a hadronic basis various contributions to $G_M^s$. In either case, one must
rely upon experimental input to determine these contributions. In practice, dispersion relations
and $\chi$PT have yet to yield entirely model-independent predictions, though considerable
insight has been derived from their application.

The benchmark dispersion
relation analysis was performed by Jaffe~\cite{Jaf89}, who employed the narrow resonance
approximation for the hadronic spectrum that is consistent with the large $N_c$ limit of
QCD and an ansatz for the high-$Q^2$ behavior of $G_M^s$ consistent with known quark
counting rules. Use of such an ansatz, which introduces a degree of
model-dependence~\cite{Forkel:1995ff},  was needed to account in a physically realistic way
for incalculable  contributions from the higher-mass part of the QCD spectrum.  The results,
which drew upon the measured isoscalar electromagnetic form factors and the flavor content of
the lowest lying $1^{--}$ mesons, suggested that $G_M^s$ would be both sizable and negative. An
updated version of this analysis was carried out by the authors of~\cite{Hammer:1995de},
who included logarithmic corrections to the power-law asymptotic behavior used in~\cite{Jaf89}
but found no substantial modification of the original Jaffe prediction.  The results in
Eq.~(\ref{eq:gmsresult}), however, tend to disfavor such a substantially negative value,
thereby suggesting a re-examination of the assumptions used in~\cite{Jaf89}.

Subsequent dispersion theoretic work in~\cite{MuH98,Ham99b,Ham99c}
avoided {\em a priori} reliance on the narrow resonance approximation and high-$Q^2$ ansatz
and instead constructed the isoscalar and strange magnetic spectral functions from measured
$K$-$N$ scattering amplitudes and $e^+e^-$ partial widths.  This analysis, which entailed a
truncation of the spectrum at $\sim$1~GeV due to the lack of experimental input, produced
results consistent with the Jaffe predictions for the low-mass spectral content of $G_M^s$ and
demonstrate that the nucleon's ``kaon cloud'' is dominated by the $\phi(1020)$ resonance.
Thus, the  narrow resonance approximation for the low-mass spectral function, as assumed
in~\cite{Jaf89}, appears justified. By itself, however, the low-mass spectral content would
imply $\mu_s\sim -0.3$, so inclusion of higher-mass contributions to the strangeness vector
spectral functions appears necessary in order to account for the SAMPLE value for $G_M^s$.
At present, no model-independent, first principles  treatment of this higher-mass region has
yet been achieved,  though a number of one-loop model
calculations~\cite{Hammer:1997vt,Gei97,Barz:1998ih,Riska:1998qd} have indicated the importance
of the higher-mass states in moving the
value $\mu_s$ in the right direction to better agree with  the SAMPLE result.

More generally, one-loop computations using a hadronic basis represent a perturbative approximation
to the full dispersion theoretic treatment. Such loop calculations entail truncation of an
expansion in the strong hadronic coupling $g$ at second order and, therefore, tend to be subject
to a substantial degree of model-dependent ambiguities~\cite{Ito:1997je,Barz:1998ih}. In particular,
the one-loop amplitudes rely on a unitarity violating approximation to pseudoscalar meson-nucleon
scattering amplitudes, the omission of higher-order (in $g$) rescattering contributions that
restore unitarity and generate the physically important resonant behavior of the spectral
functions, and use of unphysical, point-like pseudoscalar vector current form
factors~\cite{Federbush,Musolf:1996qt}. In principle, these deficiencies are remedied in a
systematic way using $\chi$PT, wherein the {\em a priori} unknown low-energy constants (LEC's)
determined by fits to experimental data embody the physics omitted from the one-loop graphs.
In the case of
$G_M^s$, however, this program encounters an intrinsic limitation due to the symmetry properties
of the strangeness vector current (see below), so any one-loop predictions for $G_M^s$
necessarily  entail unquantifiable model-dependent uncertainties.

The earliest hadron loop calculations~\cite{Koepf:hp,Musolf:1993fu,Forkel:1994yx}
suggested  that the non-resonant part of the nucleon's kaon cloud should make a fairly small
contribution to $G_M^s$, an insight confirmed by the subsequent dispersion theory analyses
discussed above.  A variation on this theme was carried out by Geiger
and Isgur~\cite{Gei97}, who summed up a complete tower of meson-baryon one loop graphs,
using  the quark model to determine the relevant hadronic vertices, and found a pattern of
cancellations  among successively higher-mass intermediate states.  Although this computation
suffers from the same ambiguities as the earlier one-loop calculations,  it is nevertheless
suggestive that  higher mass contributions may, as indicated by Eq.~(\ref{eq:gmsresult}) and
the dispersion theoretic  studies, play an important role in the dynamics of $G_M^s$.

In general, the systematic, model-independent treatment of hadronic loop effects in $\chi$PT
does not yield predictions that are independent of the $G_M^s$ measurements since the
operator ${\bar s}\gamma_\mu s$ contains  an SU(3)-singlet component~\cite{Ito:1997je}.
Apart from the SAMPLE result for  $G_M^s$ itself, there exists no other experimental
information on the SU(3)-singlet component of the nucleon vector current that would allow
one to determine the LEC's relevant to strange magnetism. An exception occurs for the strange
magnetic radius,
$<r_s^2>_M$, that governs the slope of $G_M^s$ at the origin, for which a parameter-free
prediction can be made at ${\cal O}(p^3)$~\cite{Hem98}.  However, inclusion of  ${\cal O}(p^4)$
loop contributions, nominally suppressed by one power of $m_K/\Lambda_\chi$ where
$\Lambda_\chi=4\pi F_\pi$ is the chiral scale, nearly cancels most of the ${\cal O}(p^3)$
term, leaving a residual dependence on an {\em a priori} unknown strange magnetic radius
LEC, $b_s^r$~\cite{Ham03}. Rigorously speaking, the latter must be determined from
future measurements of the $Q^2$-dependence of $G_M^s$. However, a reasonable range may
be estimated by comparing the model-independent dispersion relation and lattice QCD
calculations as indicated above, leading to $-1\leq b_s^r \leq 1$.  In this case,
$\chi$PT provides reasonable guidance for an extrapolation of $G_M^s(Q^2)$ to the
photon point.
The static moment can be written as~\cite{Ham03}
\begin{equation}
\mu_s = G_M^s(Q^2=0.1) - 0.13 b_s^r
\end{equation}
resulting in $\mu_s = 0.37\pm 0.20 \pm 0.26 \pm 0.15$, where now the two sources
of theoretical uncertainty (electroweak radiative corrections and strange magnetic radius)
have been combined.

Nucleon model calculations, though less transparently connected to QCD,  have also provided
some insights into the dynamics of strange quarks. One model that has received considerable
attention recently is the chiral quark soliton model ($\chi$QSM), in which constituent quarks
interact with the Goldstone bosons of the spontaneously-broken, approximate chiral symmetry
of QCD in a self-consistent way~\cite{Christov,Alfoker}. A recent $\chi$QSM computation of
$G_M^s$  reported in~\cite{Silva03} yields a small positive result,
$0.05\simle G_M^s(Q^2=0.1)\simle 0.1$ that is consistent with the SAMPLE result. The same
model, however, underpredicts the proton and neutron magnetic moments by $\sim 40\%$, and
the prediction for $G_M^s$ is strongly dependent on {\em ad hoc} assumptions about the
long-distance behavior of the Goldstone boson
field. The computation is nevertheless suggestive that, at the microscopic level,
the topology of the QCD vacuum plays a non-trivial role in the dynamics of the $s{\bar s}$ sea.

Completion of the SAMPLE program and the corresponding insights derived from a comparison of
the experimental results to the last decade of theoretical efforts represents
a milestone in the quest to understand
the quark substructure of the nucleon. Nonetheless, work is on-going, and
new experimental data will become available from Jefferson Laboratory and
from the Mainz PVA4 program~\cite{Maa03}. An independent determination
of $G_M^s$ at low momentum transfer is planned by the HAPPEX collaboration
using combined forward angle parity-violation measurements on hydrogen
and helium targets~\cite{Kum99,Arm00}. Precise determination of
the $Q^2$ dependence of both $G_E^s$ and $G_M^s$ over the
range $0.3<Q^2<1.0$ (GeV/c)$^2$ will be available from the planned program
of measurements by the G0 collaboration~\cite{Bec00}, for which data taking
will soon begin.  Theoretically, new efforts to carry out unquenched lattice computations and to
resolve the numerical sampling challenges are underway. One expects these efforts to flesh out the
framework that has now emerged after more than a decade of experimental and theoretical work. Of
course, future surprises are always a possibility, and one may ultimately find that some deeper
principle governs the dynamics of sea quarks~\cite{Ji:1995rd} than is apparent to our current
understanding.

This work was supported by NSF grants PHY-9420470 (Caltech),
PHY-9420787 (Illinois), PHY-9457906/9971819 (Maryland), PHY-9733772 (VPI),
DOE Cooperative agreement DE-FC02-94-ER40818 (MIT-Bates), and
DOR contract W-31-109-ENG-38 (ANL).


\begin{thebibliography}{39}

\bibitem{Kap88}
  D.~Kaplan and A.~Manohar, Nucl.~Phys.~{\bf B 310}, 527 (1988).
\bibitem{BMcK89}
  R.D.~McKeown, Phys.~Lett.~{\bf B219}, 140 (1989).
\bibitem{Bec89}
  D.H.~Beck, Phys.~Rev.~{\bf D39}, 3248 (1989).
\bibitem{Mue97} B.A.~Mueller {\it et al.}, Phys.~Rev.~Lett.~{\bf 78}, 3824 (1997).  
\bibitem{Spa00} D.T.~Spayde {\it et al.}, Phys.~Rev.~Lett.~{\bf 84}, 1106 (2000).
\bibitem{Has00} R.~Hasty {\it et al.}, Science, {\bf 290}, 2021 (2000).
\bibitem{Ani00} K.~Aniol, {\it et al.}, Phys.~Lett~{\bf B 509}, 211 (2001).
  See also K.~Aniol, {\it et al.}, Phys.~Rev.~Lett.~{\bf 82}, 1096 (1999).
\bibitem{Maa03} Mainz experiment ``PVA4'', D.~von~Harrach, spokesman.
\bibitem{Bec00} JLAB experiment E00--006 and E01--116 (``G0''), D.H.~Beck, contact.
\bibitem{Kum99} JLAB experiment E99--115, K.~Kumar and D.~Lhuillier, contacts.
\bibitem{Zhu00}
  S.-L. Zhu, S.~J.~Puglia, ~B.~R.~Holstein, and M.~J.~Ramsey-Musolf,
  Phys. Rev. D {\bf 62}, 033008 (2000).
\bibitem{Ito03} T.M.~Ito, {\it et al.}, submitted to Phys.~Rev.~Lett., nucl-ex/0310001.
\bibitem{MusR94} M.~J.~Musolf, {\it et al.}, Phys.~Rep.~{\bf 239}, 1 (1994).
\bibitem{FiJi01}
  See, for example, B.W.~Filippone and X.~Ji, Adv.~Nucl.~Phys.~{\bf 26}, 1 (2001).
\bibitem{Zel57}I.~Zel'dovich, JETP Lett.~{\bf 33}, 1531 (1957).
\bibitem{Mus90}
  M.~J.~Musolf and B.~R.~Holstein, Phys. Lett B, {\bf 242}, 461 (1990).
\bibitem{GEANT} {\sc GEANT}, CERN program library.
\bibitem{Kuc83} T.V.~Kuchto and N.M.~Shumeiko, Nucl.~Phys.~{\bf B 219}, 412 (1983).
\bibitem{Mo69} L.W.~Mo and Y.S.~Tsi, Rev.~Mod.~Phys.~{\bf 41}, 205 (1969).
\bibitem{Che01} J.-W.~Chen and X.~Ji, Phys.~Lett.~{\bf B 501}, 209 (2001).
\bibitem{Ber98} J.C.~Bergstrom, Phys.~Rev.~{\bf C58},2574 (1998).
\bibitem{Kor99} E.~Korkmaz, {\it et al.}, Phys.~Rev.~Lett.~{\bf 83}, 3609 (1999).
\bibitem{Spa01} See also D.T.~Spayde, Ph.D. thesis, Univ.~Maryland, 
2001 (unpublished but available at http://www.physics.umd.edu/enp/theses/).
\bibitem{Gal71} S.~Galster, {\it et al.}, Nucl.~Phys.~{\bf B 32}, 221 (1971).
\bibitem{Bec01} D.H.~Beck and B.R.~Holstein, Int.~J.~Mod.~Phys.~{\bf E 10}, 1 (2001).

\bibitem{Dong:1993pk}
S.~J.~Dong and K.~F.~Liu,
Phys.\ Lett.\ B {\bf 328}, 130 (1994).


\bibitem{Don98}
  S.J.~Dong, K.F.~Liu and A.G.~Williams, Phys.~Rev.~{\bf D 58}, 074504, (1998).
\bibitem{Lew03}
  R.~Lewis, W.~Wilcox, and R.M.~Woloshyn, Phys.~Rev.~{\bf D 67}, 013003 (2003).
\bibitem{Lei00}
D.~Leinweber and A.W.~Thomas, Phys.~Rev.~{\bf D 62}, 074505 (2000).

\bibitem{Jaf89}
R.L.~Jaffe, Phys.~Lett.~{\bf B229}, 275 (1989).

\bibitem{Forkel:1995ff}
H.~Forkel,
Phys.\ Rev.\ C {\bf 56}, 510 (1997).

\bibitem{Hammer:1995de}
H.~W.~Hammer, U.~G.~Meissner and D.~Drechsel,
Phys.\ Lett.\ B {\bf 367}, 323 (1996).

\bibitem{MuH98}
M.J.~Ramsey-Musolf and  H.-W.~Hammer, Phys.~Rev.~Lett.~{\bf 80}, 2539 (1998).

\bibitem{Ham99b}
H.-W.~Hammer and M.J.~Ramsey-Musolf, Phys.~Rev.~{\bf C60}, 045204 (1999) 
[Erratum-ibid. C {\bf 62}, 049902 (2000)].

\bibitem{Ham99c}
H.-W.~Hammer and M.J.~Ramsey-Musolf, 
Phys.~Rev.~{\bf C60}, 045205 (1999) [Erratum-ibid. C {\bf 62}, 049903 (2000)].

\bibitem{Hammer:1997vt}
H.~W.~Hammer and M.~J.~Ramsey-Musolf,
Phys.\ Lett.\ B {\bf 416}, 5 (1998).

\bibitem{Barz:1998ih}
L.~L.~Barz, H.~Forkel, H.~W.~Hammer, F.~S.~Navarra, M.~Nielsen and M.~J.~Ramsey-Musolf,
Nucl.\ Phys.\ A {\bf 640}, 259 (1998).

 \bibitem{Gei97}
P.~Geiger and N.~Isgur, Phys.~Rev.~{\bf C55}, 299 (1997).

\bibitem{Riska:1998qd}
D.~O.~Riska,
Few Body Syst.\ Suppl.\  {\bf 10}, 415 (1999).

\bibitem{Ito:1997je}
H.~Ito and M.~J.~Ramsey-Musolf,
Phys.\ Rev.\ C {\bf 58}, 2595 (1998).

\bibitem{Hem98}T.R.~Hemmert, U-G.~Meissner, and S.~Stieninger, Phys.~Lett.~{\bf B437}, 184 (1998).


\bibitem{Ham03} H.-W.~Hammer, S.J.~Puglia, M.J.~Ramsey-Musolf, and S.-L.~Zhu,
Phys.~Lett.~{\bf B 562}, 208 (2003).
\bibitem{Koepf:hp}
W.~Koepf, E.~M.~Henley and S.~J.~Pollock,
Phys.\ Lett.\ B {\bf 288}, 11 (1992).

\bibitem{Musolf:1993fu}
M.~J.~Musolf and M.~Burkardt,
Z.\ Phys.\ C {\bf 61}, 433 (1994).

\bibitem{Forkel:1994yx}
H.~Forkel, M.~Nielsen, X.~m.~Jin and T.~D.~Cohen,
Phys.\ Rev.\ C {\bf 50}, 3108 (1994).

\bibitem{Federbush} P. Federbush, M.L. Goldberger, and S.B. Treiman, Phys. Rev. {\bf 112}, 642 (1958).

\bibitem{Musolf:1996qt}
M.~J.~Musolf, H.~W.~Hammer and D.~Drechsel,
Phys.\ Rev.\ D {\bf 55}, 2741 (1997)
[Erratum-ibid.\ D {\bf 62}, 079901 (2000)].

\bibitem{Christov} C.V. Christov, {\em et al.}, Prog. Part. Nucl. Phys. {\bf 37}, 91 (1996).

\bibitem{Alfoker} R. Alfoker, H. Reinhardt, and H. Weigel, Phys. Rep. {\bf 265}, 139 (1996).

\bibitem{Silva03} A. Silva, H-C Kim, and K. Goeke, arXiv:hep-ph/0210189.

\bibitem{Arm00} JLab experiment E00--114, D.~Armstrong and R.~Michaels, contacts.

\bibitem{Ji:1995rd}
X.~Ji and J.~Tang,
Phys.\ Lett.\ B {\bf 362}, 182 (1995).


\end{thebibliography}
\end{document}